\documentclass[12pt,a4paper]{article}
\usepackage{amsmath}
\usepackage{xspace}
\usepackage{epsfig}
\usepackage{rotating}
\usepackage{dcolumn}
\usepackage{url}
\usepackage{hyperref}
\hypersetup{ pdfborder=0 0 0, urlbordercolor=1 0 0 }
\usepackage{cite}

\newcommand{\citenumfont}{}
\newcommand{\etal}{{\it et al.}}

\usepackage{graphicx}
\usepackage{color}
\usepackage{colortbl}
\begin{document}

%==================================================================================
%  Title page

\begin{titlepage}
%Primary authors: Sheldon Stone, Phillip Urquijo and Liming Zhang
\belowpdfbookmark{Title page}{title}

\pagenumbering{roman}
\vspace*{-1.5cm}
\centerline{\large EUROPEAN ORGANIZATION FOR NUCLEAR RESEARCH (CERN)}
\vspace*{1.5cm}
\hspace*{-5mm}\begin{tabular*}{16cm}{lc@{\extracolsep{\fill}}r}
\vspace*{-12mm}\mbox{\!\!\!\epsfig{figure=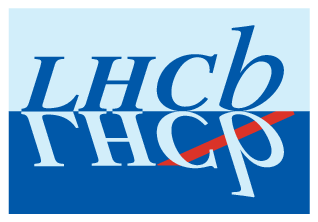,width=.12\textwidth}}& & \\
&& CERN-PH-EP-2010-029 \\
&& 14 September 2010 \\
\end{tabular*}
\vspace*{4cm}
\begin{center}
{\bf\huge\boldmath {Measurement of $\sigma\!\left(pp\to b\overline{b}X\right)$ at $\sqrt{s}=7$ TeV in the forward region}\\
}
\vspace*{2cm}
\normalsize {
The LHCb Collaboration
\footnote{Authors are listed on the following pages.}
%========================================================================%
}
\end{center}
\vspace{\fill}
\centerline{\bf Abstract}
\vspace*{5mm}\noindent
Decays of $b$ hadrons into final states containing a $D^0$ meson and a muon are used  to
measure the $b\overline{b}$ production cross-section in proton-proton
collisions at a centre-of-mass energy of 7 TeV at the LHC. In the pseudorapidity interval $2< \eta < 6$ and integrated over all transverse momenta we find that the average cross-section to produce $b$-flavoured or $\overline{b}$-flavoured hadrons is
(75.3$\pm$5.4$\pm$13.0)~$\mu$b.

\vspace*{1.cm}
\noindent{\it Keywords:} LHC, $b$-hadron,  cross-section,  bottom production\\
{\it PACS:} 14.65.Fy,  13.10.He,  13.75Cs,  13.85-t\\
\vspace{\fill}

\end{titlepage}
%\newpage
\setcounter{page}{2}
%\mbox{~}
%\newpage

\belowpdfbookmark{LHCb author list}{authors}
%\documentclass[a4paper]{article}
%\setlength{\oddsidemargin}{0cm}
%\setlength{\evensidemargin}{0cm}
%\setlength{\textwidth}{16.5cm}
%\setlength{\parindent}{0cm}
%\begin{document}
\centerline{\large\bf The LHCb Collaboration}
\begin{flushleft}
\small
%valid for 6. Aug. 2010
%to be used for data paper number 1\\
%collaborators included, who did not leave before 6. Aug. 2009\\
%                           and who joined before 6. Feb. 2010\\[2ex]
%{\small today is 14. Sep. 2010}\\[4ex]
%--
%-- LHCb Authorlist, Status of 6. Aug. 2010
%--
R.~Aaij$^{23}$,
C.~Abellan~Beteta$^{35,m}$,
B.~Adeva$^{36}$,
M.~Adinolfi$^{42}$,
C.~Adrover$^{6}$,
A.~Affolder$^{48}$,
M.~Agari$^{10}$,
Z.~Ajaltouni$^{5}$,
J.~Albrecht$^{37}$,
F.~Alessio$^{6,37}$,
M.~Alexander$^{47}$,
M.~Alfonsi$^{18}$,
P.~Alvarez~Cartelle$^{36}$,
A.A.~Alves~Jr$^{22}$,
S.~Amato$^{2}$,
Y.~Amhis$^{38}$,
J.~Amoraal$^{23}$,
J.~Anderson$^{39}$,
R.~Antunes~Nobrega$^{22,k}$,
R.~Appleby$^{50}$,
O.~Aquines~Gutierrez$^{10}$,
A.~Arefyev$^{30}$,
L.~Arrabito$^{53}$,
M.~Artuso$^{52}$,
E.~Aslanides$^{6}$,
G.~Auriemma$^{22,l}$,
S.~Bachmann$^{11}$,
Y.~Bagaturia$^{11}$,
D.S.~Bailey$^{50}$,
V.~Balagura$^{30,37}$,
W.~Baldini$^{16}$,
G.~Barber$^{49}$,
C.~Barham$^{43}$,
R.J.~Barlow$^{50}$,
S.~Barsuk$^{7}$,
S.~Basiladze$^{31}$,
A.~Bates$^{47}$,
C.~Bauer$^{10}$,
Th.~Bauer$^{23}$,
A.~Bay$^{38}$,
I.~Bediaga$^{1}$,
T.~Bellunato$^{20,i}$,
K.~Belous$^{34}$,
I.~Belyaev$^{23,30}$,
M.~Benayoun$^{8}$,
G.~Bencivenni$^{18}$,
R.~Bernet$^{39}$,
R.P.~Bernhard$^{39}$,
M.-O.~Bettler$^{17,37}$,
M.~van~Beuzekom$^{23}$,
J.H.~Bibby$^{51}$,
S.~Bifani$^{12}$,
A.~Bizzeti$^{17,g}$,
P.M.~Bj\o rnstad$^{50}$,
T.~Blake$^{49}$,
F.~Blanc$^{38}$,
C.~Blanks$^{49}$,
J.~Blouw$^{11}$,
S.~Blusk$^{52}$,
A.~Bobrov$^{33}$,
V.~Bocci$^{22}$,
B.~Bochin$^{29}$,
E.~Bonaccorsi$^{37}$,
A.~Bondar$^{33}$,
N.~Bondar$^{29,37}$,
W.~Bonivento$^{15}$,
S.~Borghi$^{47}$,
A.~Borgia$^{52}$,
E.~Bos$^{23}$,
T.J.V.~Bowcock$^{48}$,
C.~Bozzi$^{16}$,
T.~Brambach$^{9}$,
J.~van~den~Brand$^{24}$,
L.~Brarda$^{37}$,
J.~Bressieux$^{38}$,
S.~Brisbane$^{51}$,
M.~Britsch$^{10}$,
N.H.~Brook$^{42}$,
H.~Brown$^{48}$,
S.~Brusa$^{16}$,
A.~B\"{u}chler-Germann$^{39}$,
A.~Bursche$^{39}$,
J.~Buytaert$^{37}$,
S.~Cadeddu$^{15}$,
J.M.~Caicedo~Carvajal$^{37}$,
O.~Callot$^{7}$,
M.~Calvi$^{20,i}$,
M.~Calvo~Gomez$^{35,m}$,
A.~Camboni$^{35}$,
W.~Cameron$^{49}$,
L.~Camilleri$^{37}$,
P.~Campana$^{18}$,
A.~Carbone$^{14}$,
G.~Carboni$^{21,j}$,
R.~Cardinale$^{19,h}$,
A.~Cardini$^{15}$,
J.~Carroll$^{48}$,
L.~Carson$^{36}$,
K.~Carvalho~Akiba$^{23}$,
G.~Casse$^{48}$,
M.~Cattaneo$^{37}$,
B.~Chadaj$^{37}$,
M.~Charles$^{51}$,
Ph.~Charpentier$^{37}$,
J.~Cheng$^{3}$,
N.~Chiapolini$^{39}$,
A.~Chlopik$^{27}$,
J.~Christiansen$^{37}$,
P.~Ciambrone$^{18}$,
X.~Cid~Vidal$^{36}$,
P.J.~Clark$^{46}$,
P.E.L.~Clarke$^{46}$,
M.~Clemencic$^{37}$,
H.V.~Cliff$^{43}$,
J.~Closier$^{37}$,
C.~Coca$^{28}$,
V.~Coco$^{52}$,
J.~Cogan$^{6}$,
P.~Collins$^{37}$,
A.~Comerma-Montells$^{35}$,
F.~Constantin$^{28}$,
G.~Conti$^{38}$,
A.~Contu$^{51}$,
P.~Cooke$^{48}$,
M.~Coombes$^{42}$,
B.~Corajod$^{37}$,
G.~Corti$^{37}$,
G.A.~Cowan$^{46}$,
R.~Currie$^{46}$,
B.~D'Almagne$^{7}$,
C.~D'Ambrosio$^{37}$,
I.~D'Antone$^{14}$,
W.~Da~Silva$^{8}$,
E.~Dane'$^{18}$,
P.~David$^{8}$,
I.~De~Bonis$^{4}$,
S.~De~Capua$^{21,j}$,
M.~De~Cian$^{39}$,
F.~De~Lorenzi$^{12}$,
J.M.~De~Miranda$^{1}$,
L.~De~Paula$^{2}$,
P.~De~Simone$^{18}$,
D.~Decamp$^{4}$,
G.~Decreuse$^{37}$,
H.~Degaudenzi$^{38,37}$,
M.~Deissenroth$^{11}$,
L.~Del~Buono$^{8}$,
C.J.~Densham$^{45}$,
C.~Deplano$^{15}$,
O.~Deschamps$^{5}$,
F.~Dettori$^{15,c}$,
J.~Dickens$^{43}$,
H.~Dijkstra$^{37}$,
M.~Dima$^{28}$,
S.~Donleavy$^{48}$,
P.~Dornan$^{49}$,
D.~Dossett$^{44}$,
A.~Dovbnya$^{40}$,
R.~Dumps$^{37}$,
F.~Dupertuis$^{38}$,
L.~Dwyer$^{48}$,
R.~Dzhelyadin$^{34}$,
C.~Eames$^{49}$,
S.~Easo$^{45}$,
U.~Egede$^{49}$,
V.~Egorychev$^{30}$,
S.~Eidelman$^{33}$,
D.~van~Eijk$^{23}$,
F.~Eisele$^{11}$,
S.~Eisenhardt$^{46}$,
L.~Eklund$^{47}$,
D.G.~d'Enterria$^{35,n}$,
D.~Esperante~Pereira$^{36}$,
L.~Est\`{e}ve$^{43}$,
E.~Fanchini$^{20,i}$,
C.~F\"{a}rber$^{11}$,
G.~Fardell$^{46}$,
C.~Farinelli$^{23}$,
S.~Farry$^{12}$,
V.~Fave$^{38}$,
G.~Felici$^{18}$,
V.~Fernandez~Albor$^{36}$,
M.~Ferro-Luzzi$^{37}$,
S.~Filippov$^{32}$,
C.~Fitzpatrick$^{46}$,
W.~Flegel$^{37}$,
F.~Fontanelli$^{19,h}$,
C.~Forti$^{18}$,
R.~Forty$^{37}$,
C.~Fournier$^{37}$,
B.~Franek$^{45}$,
M.~Frank$^{37}$,
C.~Frei$^{37}$,
M.~Frosini$^{17,e}$,
J.L.~Fungueirino~Pazos$^{36}$,
S.~Furcas$^{20}$,
A.~Gallas~Torreira$^{36}$,
D.~Galli$^{14,b}$,
M.~Gandelman$^{2}$,
P.~Gandini$^{51}$,
Y.~Gao$^{3}$,
J-C.~Garnier$^{37}$,
L.~Garrido$^{35}$,
D.~Gascon$^{35}$,
C.~Gaspar$^{37}$,
A.~Gaspar~De~Valenzuela~Cue$^{35,m}$,
J.~Gassner$^{39}$,
N.~Gauvin$^{38}$,
P.~Gavillet$^{37}$,
M.~Gersabeck$^{37}$,
T.~Gershon$^{44}$,
Ph.~Ghez$^{4}$,
V.~Gibson$^{43}$,
V.V.~Gligorov$^{37}$,
C.~G\"{o}bel$^{54}$,
D.~Golubkov$^{30}$,
A.~Golutvin$^{49,30,37}$,
A.~Gomes$^{1}$,
G.~Gong$^{3}$,
H.~Gong$^{3}$,
H.~Gordon$^{51}$,
M.~Grabalosa~G\'{a}ndara$^{35}$,
V.~Gracco$^{19,h}$,
R.~Graciani~Diaz$^{35}$,
L.A.~Granado~Cardoso$^{37}$,
E.~Graug\'{e}s$^{35}$,
G.~Graziani$^{17}$,
A.~Grecu$^{28}$,
S.~Gregson$^{43}$,
G.~Guerrer$^{1}$,
B.~Gui$^{52}$,
E.~Gushchin$^{32}$,
Yu.~Guz$^{34,37}$,
Z.~Guzik$^{27}$,
T.~Gys$^{37}$,
G.~Haefeli$^{38}$,
S.C.~Haines$^{43}$,
T.~Hampson$^{42}$,
S.~Hansmann-Menzemer$^{11}$,
R.~Harji$^{49}$,
N.~Harnew$^{51}$,
P.F.~Harrison$^{44}$,
J.~He$^{7}$,
K.~Hennessy$^{48}$,
P.~Henrard$^{5}$,
J.A.~Hernando~Morata$^{36}$,
E.~van~Herwijnen$^{37}$,
A.~Hicheur$^{38}$,
E.~Hicks$^{48}$,
H.J.~Hilke$^{37}$,
W.~Hofmann$^{10}$,
K.~Holubyev$^{11}$,
P.~Hopchev$^{4}$,
W.~Hulsbergen$^{23}$,
P.~Hunt$^{51}$,
T.~Huse$^{48}$,
R.S.~Huston$^{12}$,
D.~Hutchcroft$^{48}$,
F.~Iacoangeli$^{22}$,
V.~Iakovenko$^{7,41}$,
C.~Iglesias~Escudero$^{36}$,
C.~Ilgner$^{9}$,
P.~Ilten$^{12}$,
J.~Imong$^{42}$,
R.~Jacobsson$^{37}$,
M.~Jahjah~Hussein$^{5}$,
O.~Jamet$^{37}$,
E.~Jans$^{23}$,
F.~Jansen$^{23}$,
P.~Jaton$^{38}$,
B.~Jean-Marie$^{7}$,
M.~John$^{51}$,
D.~Johnson$^{51}$,
C.R.~Jones$^{43}$,
B.~Jost$^{37}$,
F.~Kapusta$^{8}$,
T.M.~Karbach$^{9}$,
A.~Kashchuk$^{29}$,
S.~Katvars$^{43}$,
J.~Keaveney$^{12}$,
U.~Kerzel$^{43}$,
T.~Ketel$^{24}$,
A.~Keune$^{38}$,
S.~Khalil$^{52}$,
B.~Khanji$^{6}$,
Y.M.~Kim$^{46}$,
M.~Knecht$^{38}$,
S.~Koblitz$^{37}$,
A.~Konoplyannikov$^{30}$,
P.~Koppenburg$^{23}$,
M.~Korolev$^{31}$,
A.~Kozlinskiy$^{23}$,
L.~Kravchuk$^{32}$,
R.~Kristic$^{37}$,
G.~Krocker$^{11}$,
P.~Krokovny$^{11}$,
F.~Kruse$^{9}$,
K.~Kruzelecki$^{37}$,
M.~Kucharczyk$^{25}$,
I.~Kudryashov$^{31}$,
S.~Kukulak$^{25}$,
R.~Kumar$^{14,37}$,
T.~Kvaratskheliya$^{30}$,
V.N.~La~Thi$^{38}$,
D.~Lacarrere$^{37}$,
G.~Lafferty$^{50}$,
A.~Lai$^{15}$,
R.W.~Lambert$^{37}$,
G.~Lanfranchi$^{18}$,
C.~Langenbruch$^{11}$,
T.~Latham$^{44}$,
R.~Le~Gac$^{6}$,
J.-P.~Lees$^{4}$,
R.~Lef\`{e}vre$^{5}$,
A.~Leflat$^{31,37}$,
J.~Lefran\c{c}ois$^{7}$,
F.~Lehner$^{39}$,
M.~Lenzi$^{17}$,
O.~Leroy$^{6}$,
T.~Lesiak$^{25}$,
L.~Li$^{3}$,
Y.Y.~Li$^{43}$,
L.~Li~Gioi$^{5}$,
J.~Libby$^{51}$,
M.~Lieng$^{9}$,
R.~Lindner$^{37}$,
S.~Lindsay$^{48}$,
C.~Linn$^{11}$,
B.~Liu$^{3}$,
G.~Liu$^{37}$,
S.~L\"{o}chner$^{10}$,
J.H.~Lopes$^{2}$,
E.~Lopez~Asamar$^{35}$,
N.~Lopez-March$^{38}$,
P.~Loveridge$^{45}$,
J.~Luisier$^{38}$,
B.~M'charek$^{24}$,
F.~Machefert$^{7}$,
I.V.~Machikhiliyan$^{4,30}$,
F.~Maciuc$^{10}$,
O.~Maev$^{29}$,
J.~Magnin$^{1}$,
A.~Maier$^{37}$,
S.~Malde$^{51}$,
R.M.D.~Mamunur$^{37}$,
G.~Manca$^{15,c,37}$,
G.~Mancinelli$^{6}$,
N.~Mangiafave$^{43}$,
U.~Marconi$^{14}$,
R.~M\"{a}rki$^{38}$,
J.~Marks$^{11}$,
G.~Martellotti$^{22}$,
A.~Martens$^{7}$,
L.~Martin$^{51}$,
D.~Martinez~Santos$^{36}$,
A.~Massafferri$^{1}$,
Z.~Mathe$^{12}$,
C.~Matteuzzi$^{20}$,
V.~Matveev$^{34}$,
E.~Maurice$^{6}$,
B.~Maynard$^{52}$,
A.~Mazurov$^{32}$,
G.~McGregor$^{50}$,
R.~McNulty$^{12}$,
C.~Mclean$^{14}$,
M.~Merk$^{23}$,
J.~Merkel$^{9}$,
M.~Merkin$^{31}$,
R.~Messi$^{21,j}$,
S.~Miglioranzi$^{37}$,
M.-N.~Minard$^{4}$,
G.~Moine$^{37}$,
S.~Monteil$^{5}$,
D.~Moran$^{12}$,
J.~Morant$^{37}$,
P.~Morawski$^{25}$,
J.V.~Morris$^{45}$,
J.~Moscicki$^{37}$,
R.~Mountain$^{52}$,
I.~Mous$^{23}$,
F.~Muheim$^{46}$,
K.~M\"{u}ller$^{39}$,
R.~Muresan$^{38}$,
F.~Murtas$^{18}$,
B.~Muryn$^{26}$,
M.~Musy$^{35}$,
J.~Mylroie-Smith$^{48}$,
P.~Naik$^{42}$,
T.~Nakada$^{38}$,
R.~Nandakumar$^{45}$,
J.~Nardulli$^{45}$,
A.~Nawrot$^{27}$,
M.~Nedos$^{9}$,
M.~Needham$^{38}$,
N.~Neufeld$^{37}$,
P.~Neustroev$^{29}$,
M.~Nicol$^{7}$,
L.~Nicolas$^{38}$,
S.~Nies$^{9}$,
V.~Niess$^{5}$,
N.~Nikitin$^{31}$,
A.~Noor$^{48}$,
A.~Oblakowska-Mucha$^{26}$,
V.~Obraztsov$^{34}$,
S.~Oggero$^{23}$,
O.~Okhrimenko$^{41}$,
R.~Oldeman$^{15,c}$,
M.~Orlandea$^{28}$,
A.~Ostankov$^{34}$,
B.~Pal$^{52}$,
J.~Palacios$^{39}$,
M.~Palutan$^{18}$,
J.~Panman$^{37}$,
A.~Papadelis$^{23}$,
A.~Papanestis$^{45}$,
M.~Pappagallo$^{13,a}$,
C.~Parkes$^{47,37}$,
C.J.~Parkinson$^{49}$,
G.~Passaleva$^{17}$,
G.D.~Patel$^{48}$,
M.~Patel$^{49}$,
S.K.~Paterson$^{49,37}$,
G.N.~Patrick$^{45}$,
C.~Patrignani$^{19,h}$,
E.~Pauna$^{28}$,
C.~Pauna~(Chiojdeanu)$^{28}$,
C.~Pavel~(Nicorescu)$^{28}$,
A.~Pazos~Alvarez$^{36}$,
A.~Pellegrino$^{23}$,
G.~Penso$^{22,k}$,
M.~Pepe~Altarelli$^{37}$,
S.~Perazzini$^{14,b}$,
D.L.~Perego$^{20,i}$,
E.~Perez~Trigo$^{36}$,
A.~P\'{e}rez-Calero~Yzquierdo$^{35}$,
P.~Perret$^{5}$,
G.~Pessina$^{20}$,
A.~Petrella$^{16,d,37}$,
A.~Petrolini$^{19,h}$,
E.~Picatoste~Olloqui$^{35}$,
B.~Pie~Valls$^{35}$,
D.~Piedigrossi$^{37}$,
B.~Pietrzyk$^{4}$,
D.~Pinci$^{22}$,
S.~Playfer$^{46}$,
M.~Plo~Casasus$^{36}$,
M.~Poli-Lener$^{18}$,
G.~Polok$^{25}$,
A.~Poluektov$^{44,33}$,
E.~Polycarpo$^{2}$,
D.~Popov$^{10}$,
B.~Popovici$^{28}$,
S.~Poss$^{6}$,
C.~Potterat$^{38}$,
A.~Powell$^{51}$,
S.~Pozzi$^{16,d}$,
T.~du~Pree$^{23}$,
V.~Pugatch$^{41}$,
A.~Puig~Navarro$^{35}$,
W.~Qian$^{3,7}$,
J.H.~Rademacker$^{42}$,
B.~Rakotomiaramanana$^{38}$,
I.~Raniuk$^{40}$,
G.~Raven$^{24}$,
S.~Redford$^{51}$,
W.~Reece$^{49}$,
A.C.~dos~Reis$^{1}$,
S.~Ricciardi$^{45}$,
J.~Riera$^{35,m}$,
K.~Rinnert$^{48}$,
D.A.~Roa~Romero$^{5}$,
P.~Robbe$^{7,37}$,
E.~Rodrigues$^{47}$,
F.~Rodrigues$^{2}$,
C.~Rodriguez~Cobo$^{36}$,
P.~Rodriguez~Perez$^{36}$,
G.J.~Rogers$^{43}$,
V.~Romanovsky$^{34}$,
E.~Rondan~Sanabria$^{1}$,
M.~Rosello$^{35,m}$,
J.~Rouvinet$^{38}$,
L.~Roy$^{37}$,
T.~Ruf$^{37}$,
H.~Ruiz$^{35}$,
C.~Rummel$^{11}$,
V.~Rusinov$^{30}$,
G.~Sabatino$^{21,j}$,
J.J.~Saborido~Silva$^{36}$,
N.~Sagidova$^{29}$,
P.~Sail$^{47}$,
B.~Saitta$^{15,c}$,
T.~Sakhelashvili$^{39}$,
C.~Salzmann$^{39}$,
A.~Sambade~Varela$^{37}$,
M.~Sannino$^{19,h}$,
R.~Santacesaria$^{22}$,
R.~Santinelli$^{37}$,
E.~Santovetti$^{21,j}$,
M.~Sapunov$^{6}$,
A.~Sarti$^{18}$,
C.~Satriano$^{22,l}$,
A.~Satta$^{21}$,
T.~Savidge$^{49}$,
M.~Savrie$^{16,d}$,
D.~Savrina$^{30}$,
P.~Schaack$^{49}$,
M.~Schiller$^{11}$,
S.~Schleich$^{9}$,
M.~Schmelling$^{10}$,
B.~Schmidt$^{37}$,
O.~Schneider$^{38}$,
T.~Schneider$^{37}$,
A.~Schopper$^{37}$,
M.-H.~Schune$^{7}$,
R.~Schwemmer$^{37}$,
A.~Sciubba$^{18,k}$,
M.~Seco$^{36}$,
A.~Semennikov$^{30}$,
K.~Senderowska$^{26}$,
N.~Serra$^{23}$,
J.~Serrano$^{6}$,
B.~Shao$^{3}$,
M.~Shapkin$^{34}$,
I.~Shapoval$^{40,37}$,
P.~Shatalov$^{30}$,
Y.~Shcheglov$^{29}$,
T.~Shears$^{48}$,
L.~Shekhtman$^{33}$,
V.~Shevchenko$^{30}$,
A.~Shires$^{49}$,
S.~Sigurdsson$^{43}$,
E.~Simioni$^{24}$,
H.P.~Skottowe$^{43}$,
T.~Skwarnicki$^{52}$,
N.~Smale$^{10,51}$,
A.~Smith$^{37}$,
A.C.~Smith$^{37}$,
N.A.~Smith$^{48}$,
K.~Sobczak$^{5}$,
F.J.P.~Soler$^{47}$,
A.~Solomin$^{42}$,
P.~Somogy$^{37}$,
F.~Soomro$^{49}$,
B.~Souza~De~Paula$^{2}$,
B.~Spaan$^{9}$,
A.~Sparkes$^{46}$,
E.~Spiridenkov$^{29}$,
P.~Spradlin$^{51}$,
A.~Srednicki$^{27}$,
F.~Stagni$^{37}$,
S.~Stahl$^{11}$,
S.~Steiner$^{39}$,
O.~Steinkamp$^{39}$,
O.~Stenyakin$^{34}$,
S.~Stoica$^{28}$,
S.~Stone$^{52}$,
B.~Storaci$^{23}$,
U.~Straumann$^{39}$,
N.~Styles$^{46}$,
M.~Szczekowski$^{27}$,
P.~Szczypka$^{38}$,
T.~Szumlak$^{47,26}$,
S.~T'Jampens$^{4}$,
E.~Tarkovskiy$^{30}$,
E.~Teodorescu$^{28}$,
H.~Terrier$^{23}$,
F.~Teubert$^{37}$,
C.~Thomas$^{51,45}$,
E.~Thomas$^{37}$,
J.~van~Tilburg$^{39}$,
V.~Tisserand$^{4}$,
M.~Tobin$^{39}$,
S.~Topp-Joergensen$^{51}$,
M.T.~Tran$^{38}$,
S.~Traynor$^{12}$,
U.~Trunk$^{10}$,
A.~Tsaregorodtsev$^{6}$,
N.~Tuning$^{23}$,
A.~Ukleja$^{27}$,
O.~Ullaland$^{37}$,
P.~Urquijo$^{52}$,
U.~Uwer$^{11}$,
V.~Vagnoni$^{14}$,
G.~Valenti$^{14}$,
R.~Vazquez~Gomez$^{35}$,
P.~Vazquez~Regueiro$^{36}$,
S.~Vecchi$^{16}$,
J.J.~Velthuis$^{42}$,
M.~Veltri$^{17,f}$,
K.~Vervink$^{37}$,
B.~Viaud$^{7}$,
I.~Videau$^{7}$,
X.~Vilasis-Cardona$^{35,m}$,
J.~Visniakov$^{36}$,
A.~Vollhardt$^{39}$,
D.~Volyanskyy$^{39}$,
D.~Voong$^{42}$,
A.~Vorobyev$^{29}$,
An.~Vorobyev$^{29}$,
H.~Voss$^{10}$,
K.~Wacker$^{9}$,
S.~Wandernoth$^{11}$,
J.~Wang$^{52}$,
D.R.~Ward$^{43}$,
A.D.~Webber$^{50}$,
D.~Websdale$^{49}$,
M.~Whitehead$^{44}$,
D.~Wiedner$^{11}$,
L.~Wiggers$^{23}$,
G.~Wilkinson$^{51}$,
M.P.~Williams$^{44}$,
M.~Williams$^{49}$,
F.F.~Wilson$^{45}$,
J.~Wishahi$^{9}$,
M.~Witek$^{25}$,
W.~Witzeling$^{37}$,
M.L.~Woodward$^{45}$,
S.A.~Wotton$^{43}$,
K.~Wyllie$^{37}$,
Y.~Xie$^{46}$,
F.~Xing$^{51}$,
Z.~Yang$^{3}$,
G.~Ybeles~Smit$^{23}$,
R.~Young$^{46}$,
O.~Yushchenko$^{34}$,
M.~Zeng$^{3}$,
L.~Zhang$^{52}$,
Y.~Zhang$^{3}$,
A.~Zhelezov$^{11}$,
E.~Zverev$^{31}$.\bigskip\newline{\it
\footnotesize
$ ^{1}$Centro Brasileiro de Pesquisas F\'{i}sicas (CBPF), Rio de Janeiro, Brazil\\
$ ^{2}$Universidade Federal do Rio de Janeiro (UFRJ), Rio de Janeiro, Brazil\\
$ ^{3}$Center for High Energy Physics, Tsinghua University, Beijing, China\\
$ ^{4}$LAPP, Universit\'{e} de Savoie, CNRS/IN2P3, Annecy-Le-Vieux, France\\
$ ^{5}$Clermont Universit\'{e}, Universit\'{e} Blaise Pascal, CNRS/IN2P3, LPC, Clermont-Ferrand, France\\
$ ^{6}$CPPM, Aix-Marseille Universit\'{e}, CNRS/IN2P3, Marseille, France\\
$ ^{7}$LAL, Universit\'{e} Paris-Sud, CNRS/IN2P3, Orsay, France\\
$ ^{8}$LPNHE, Universit\'{e} Pierre et Marie Curie, Universit\'{e} Paris Diderot, CNRS/IN2P3, Paris, France\\
$ ^{9}$Fakult\"{a}t Physik, Technische Universit\"{a}t Dortmund, Dortmund, Germany\\
$ ^{10}$Max-Planck-Institut f\"{u}r Kernphysik (MPIK), Heidelberg, Germany\\
$ ^{11}$Physikalisches Institut, Ruprecht-Karls-Universit\"{a}t Heidelberg, Heidelberg, Germany\\
$ ^{12}$School of Physics, University College Dublin, Dublin, Ireland\\
$ ^{13}$Sezione INFN di Bari, Bari, Italy\\
$ ^{14}$Sezione INFN di Bologna, Bologna, Italy\\
$ ^{15}$Sezione INFN di Cagliari, Cagliari, Italy\\
$ ^{16}$Sezione INFN di Ferrara, Ferrara, Italy\\
$ ^{17}$Sezione INFN di Firenze, Firenze, Italy\\
$ ^{18}$Laboratori Nazionali dell'INFN di Frascati, Frascati, Italy\\
$ ^{19}$Sezione INFN di Genova, Genova, Italy\\
$ ^{20}$Sezione INFN di Milano Bicocca, Milano, Italy\\
$ ^{21}$Sezione INFN di Roma Tor Vergata, Roma, Italy\\
$ ^{22}$Sezione INFN di Roma Sapienza, Roma, Italy\\
$ ^{23}$Nikhef National Institute for Subatomic Physics, Amsterdam, Netherlands\\
$ ^{24}$Nikhef National Institute for Subatomic Physics and Vrije Universiteit, Amsterdam, Netherlands\\
$ ^{25}$Henryk Niewodniczanski Institute of Nuclear Physics, Polish Academy of Sciences, Cracow, Poland\\
$ ^{26}$Faculty of Physics \& Applied Computer Science, Cracow, Poland\\
$ ^{27}$Soltan Institute for Nuclear Studies, Warsaw, Poland\\
$ ^{28}$Horia Hulubei National Institute of Physics and Nuclear Engineering, Bucharest-Magurele, Romania\\
$ ^{29}$Petersburg Nuclear Physics Institute (PNPI), Gatchina, Russia\\
$ ^{30}$Institute of Theoretical and Experimental Physics (ITEP), Moscow, Russia\\
$ ^{31}$Institute of Nuclear Physics, Moscow State University (SINP MSU), Moscow, Russia\\
$ ^{32}$Institute for Nuclear Research of the Russian Academy of Sciences (INR RAN), Moscow, Russia\\
$ ^{33}$Budker Institute of Nuclear Physics (BINP), Novosibirsk, Russia\\
$ ^{34}$Institute for High Energy Physics (IHEP), Protvino, Russia\\
$ ^{35}$Universitat de Barcelona, Barcelona, Spain\\
$ ^{36}$Universidad de Santiago de Compostela, Santiago de Compostela, Spain\\
$ ^{37}$European Organization for Nuclear Research (CERN), Geneva, Switzerland\\
$ ^{38}$Ecole Polytechnique F\'{e}d\'{e}rale de Lausanne (EPFL), Lausanne, Switzerland\\
$ ^{39}$Physik-Institut, Universit\"{a}t Z\"{u}rich, Z\"{u}rich, Switzerland\\
$ ^{40}$NSC Kharkiv Institute of Physics and Technology (NSC KIPT), Kharkiv, Ukraine\\
$ ^{41}$Institute for Nuclear Research of the National Academy of Sciences (KINR), Kyiv, Ukraine\\
$ ^{42}$H.H. Wills Physics Laboratory, University of Bristol, Bristol, United Kingdom\\
$ ^{43}$Cavendish Laboratory, University of Cambridge, Cambridge, United Kingdom\\
$ ^{44}$Department of Physics, University of Warwick, Coventry, United Kingdom\\
$ ^{45}$STFC Rutherford Appleton Laboratory, Didcot, United Kingdom\\
$ ^{46}$School of Physics and Astronomy, University of Edinburgh, Edinburgh, United Kingdom\\
$ ^{47}$School of Physics and Astronomy, University of Glasgow, Glasgow, United Kingdom\\
$ ^{48}$Oliver Lodge Laboratory, University of Liverpool, Liverpool, United Kingdom\\
$ ^{49}$Imperial College London, London, United Kingdom\\
$ ^{50}$School of Physics and Astronomy, University of Manchester, Manchester, United Kingdom\\
$ ^{51}$Department of Physics, University of Oxford, Oxford, United Kingdom\\
$ ^{52}$Syracuse University, Syracuse, NY, United States\\
$ ^{53}$CC-IN2P3, CNRS/IN2P3, Lyon-Villeurbanne, France, associated member\\
$ ^{54}$Pontif\'{i}cia Universidade Cat\'{o}lica do Rio de Janeiro (PUC-Rio), Rio de Janeiro, Brazil, associated to $^2 $\\
\bigskip
$ ^{a}$Universit\`{a} di Bari, Bari, Italy\\
$ ^{b}$Universit\`{a} di Bologna, Bologna, Italy\\
$ ^{c}$Universit\`{a} di Cagliari, Cagliari, Italy\\
$ ^{d}$Universit\`{a} di Ferrara, Ferrara, Italy\\
$ ^{e}$Universit\`{a} di Firenze, Firenze, Italy\\
$ ^{f}$Universit\`{a} di Urbino, Urbino, Italy\\
$ ^{g}$Universit\`{a} di Modena e Reggio Emilia, Modena, Italy\\
$ ^{h}$Universit\`{a} di Genova, Genova, Italy\\
$ ^{i}$Universit\`{a} di Milano Bicocca, Milano, Italy\\
$ ^{j}$Universit\`{a} di Roma Tor Vergata, Roma, Italy\\
$ ^{k}$Universit\`{a} di Roma La Sapienza, Roma, Italy\\
$ ^{l}$Universit\`{a} della Basilicata, Potenza, Italy\\
$ ^{m}$LIFAELS, La Salle, Universitat Ramon Llull, Barcelona, Spain\\
$ ^{n}$Instituci\'{o} Catalana de Recerca i Estudis Avan\c{c}ats (ICREA), Barcelona, Spain\\
}
%--
%-- LHCb Authorlist, Status 6. Aug. 2010
%-- Number of Authors = 630
%--
\end{flushleft}
%\end{document}

\cleardoublepage
\setcounter{page}{1}
\pagenumbering{arabic}

%==========================

%\maketitle
%\begin{keyword}
%LHC ,  $b$-hadron ,  cross-section ,  bottom production
%\PACS 14.65.Fy ,  13.10.He ,  13.75Cs ,  13.85-t
%\end{keyword}
% end{frontmatter}

\newpage
\section{Introduction}
\noindent Quantum Chromodynamics predicts the cross-section for the production of $b$-flavoured hadrons in proton-proton collisions, for which higher order calculations are available \cite{prod-theory}. The first data taken with the LHCb experiment at 7 TeV centre-of-mass energy allows this cross-section to be measured and compared to predictions. Knowledge of the $b$ yield is also critical in ascertaining the sensitivity of experiments that aim to measure fundamental parameters of interest involving, for example, $CP$ violation.  It is also useful for normalising backgrounds for measurements of higher mass objects that decay into $b\overline{b}$, such as the Higgs boson. In this paper we present a measurement of the
production cross-section for the average of  $b$-flavoured and $\overline{b}$-flavoured hadrons
in proton-proton collisions at a centre-of-mass energy of 7 TeV in the pseudorapidity interval $2< \eta < 6$, where $\eta=-{\rm ln}\left[\tan(\theta/2)\right]$,
and $\theta$ is the angle of the weakly decaying $b$ or $\overline{b}$ hadron with respect to the proton direction. We extrapolate this measurement to the entire rapidity interval. Our sensitivity extends over the entire range of transverse momentum of the $b$-flavoured hadron.

The LHCb detector \cite{LHCb-det} was constructed as a forward spectrometer primarily to measure $CP$ violating and rare decays of hadrons containing $b$ and $c$ quarks.
The detector elements are placed along the beam line of the LHC starting with the Vertex Locator (VELO), a silicon strip device that surrounds the proton-proton interaction region and is positioned 8 mm from the beam during collisions. It provides precise locations for primary $pp$ interaction vertices, the locations of decays of long lived hadrons, and contributes to the measurement of track momenta.  Other detectors used to measure track momenta comprise a large area silicon strip detector located before a 3.7 Tm dipole magnet, and a combination of
silicon strip detectors and straw drift chambers placed afterward. Two Ring Imaging Cherenkov (RICH) detectors are used to identify charged hadrons. Further downstream an Electromagnetic Calorimeter (ECAL) is used for photon detection and electron identification, followed by a Hadron Calorimeter (HCAL), and  a system consisting of alternating layers of iron and chambers (MWPC and triple-GEM) that distinguishes muons from hadrons (MUON). The ECAL, MUON, and HCAL provide the capability of
first-level hardware triggering.

Two independent data samples, recorded at different times, are examined.
 For the earliest period of data taking the number of colliding bunches was
sufficiently low that the high-level trigger could process all crossings and
 accept events when at least one track was reconstructed in either the VELO or the tracking stations. This data set, called  ``microbias", has an integrated luminosity, ${\cal {L}}$, of 2.9 nb$^{-1}$. The second sample, referred to as ``triggered", uses triggers designed to select a single muon. Here ${\cal {L}}$ equals 12.2 nb$^{-1}$. These samples are analysed  independently and the results subsequently combined.

Most $D^0$ mesons are produced directly via $pp\to c\overline{c} X$ interactions, where $X$ indicates any combination of final state particles. These particular  $D^0$ mesons are denoted as ``Prompt".  $D^0$  mesons  produced in $pp\to b\overline{b} X$ collisions where the $b$-flavoured hadron decays into a final state containing a $D^0$ meson are called ``Dfb". We use the decay channel $b\to D^0 X \mu^-\overline{\nu}$, as it has a large branching fraction of  (6.84$\pm$0.35)\% \cite{Lepers}, and is advantageous from the point of view of signal to background. Throughout this paper mention of a particular mode implies the inclusion of the charge conjugate mode as well.

\section{\boldmath Analysis of $D^0\to K^-\pi^+$}

The Prompt and Dfb $D^0$ components can be separated statistically by examining the impact parameter (IP)
 with respect to the closest primary vertex, where IP is defined as the smallest distance between the $D^0$ reconstructed trajectory and the primary vertex \footnote{Primary vertices are found using an iterative procedure based on the closest approach of tracks with each other. The resolutions of the resulting vertex positions depend on the number of tracks and are of the order of 70 $\mu$m along the beam direction and 10 $\mu$m in each transverse coordinate, for 40 tracks.}. We use the $D^0\to K^-\pi^+$ channel which has a branching fraction of (3.89$\pm$0.05)\% \cite{PDG}.

The $D^0$ selection criteria are the same regardless of the trigger conditions. Both the kaon and pion candidates are associated with
Cherenkov photons in the RICH system. The photon angles with respect to the track direction are examined and a likelihood formed for each particle hypothesis \cite{LHCb-det}. Candidates are identified as kaons or pions on the basis of this likelihood.  We also require that the momentum transverse to the beam direction, $p_{\rm T}$, of both the kaon and pion be $>$ 300 MeV, and that their scalar sum is  $>$1400 MeV. (We work in units with $c$=1.)
Since real $D^0$ mesons travel before decaying, the kaon and pion tracks when followed backwards will most often not point to the closest primary vertex. We require that the $\chi^2$ formed by using the hypothesis that the impact parameter is equal to zero, $\chi^2_{\rm IP}$, be $>9$ for each track. They also must be consistent with coming from a common origin with vertex fit $\chi^2<6$.
%Finally, the $D^0$ candidate must be detached from the closest primary vertex with
 %flight distance significance $\chi^2_{\rm FS}>64$, where the flight distance is measured between the primary and $D^0$ vertices, and % $\chi^2_{\rm FS}$ is formed by using the hypothesis that the flight distance is zero.
 % Finally, the $D^0$ candidate must be detached from the closest primary vertex requiring a $\chi^2$ based on the flight-distance %significance,$\chi^2_{\rm FS}>64$, which is formed by using the hypothesis that the flight distance between the primary and $D^0$ vertices is % zero.
Finally, the $D^0$ candidate must be detached from the closest primary vertex. To implement this flight distance significance test we form a $\chi^2_{\rm FS}$ based on the hypothesis that the flight distance between the primary and $D^0$ vertices is zero, and require $\chi^2_{\rm FS}>64$.
This set of requirements on the $D^0$  candidate is labeled ``generic". All of these requirements were selected by comparing sidebands of the invariant $K^-\pi^+$ mass distribution, representative of the background, with signal Monte Carlo simulation using PYTHIA 6.4 \cite{Pythia}.

\begin{figure}[bt]
\centering
\includegraphics[width=3.5in]{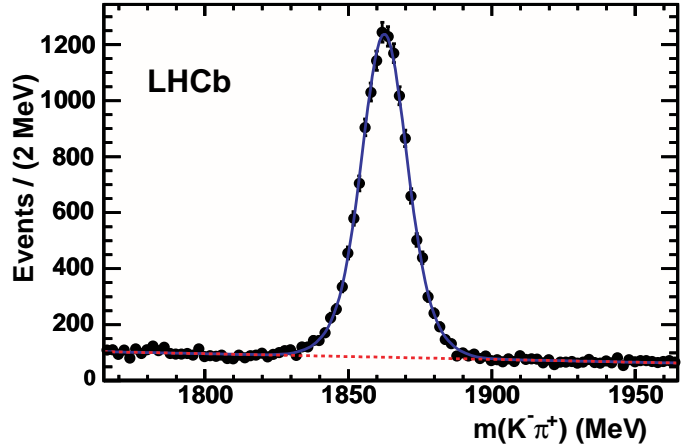}
\caption{\small $K^-\pi^+$ invariant mass for ``Prompt" selection
criteria in 2.9 nb$^{-1}$. The curve shows a fit to a linear background (dashed) plus double-Gaussian signal function with parameters $\sigma_1$=7.1$\pm$0.6 MeV, $\sigma_2/\sigma_1$=1.7$\pm$0.1, and the fraction of the second Gaussian 0.40$\pm$0.16.
 } \label{d0mass-530}
\end{figure}

In order to ascertain the parameters characterizing the $D^0$ mass peak,  a sample enriched in Prompt $D^0$  mesons is selected. This is achieved by including two additional requirements:  (1) the cosine of the angle between the $D^0$ candidate's momentum direction and the line from the $K^-\pi^+$ vertex to the primary vertex must be $>0.9999$, and (2) the $\chi^2_{\rm IP}$ for the $D^0$ must be less than 25. The $K^-\pi^+$ invariant mass distribution after imposing all of these requirements is shown in Fig.~\ref{d0mass-530}. The data are fit with a double-Gaussian signal function, with both Gaussians having the same mean, and a linear background. This signal shape is used in all subsequent fits.

\begin{figure}[hbt]
\centering
\includegraphics[width=3.in]{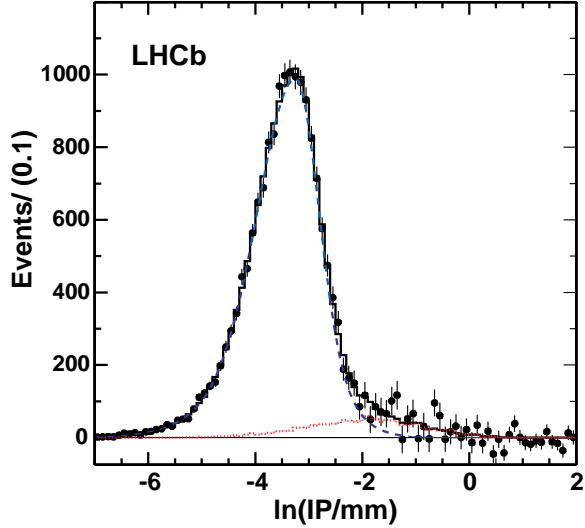}
\caption{\small Natural logarithm of the IP for $D^0$  mesons, with the IP in units of mm (points with error bars) for the 2.9 nb$^{-1}$ microbias sample. Background has been subtracted using mass sidebands.
The dashed curve shows the result of the fit to the Prompt component, the dotted line the Dfb component, and the histogram the sum of the two.}
\label{d0lnip_ss}
\end{figure}

Selecting $K^-\pi^+$ candidates within $\pm$20 MeV of the fitted $D^0$ mass peak and subtracting the background using  invariant
mass sidebands 35-75 MeV from the peak on both sides, we display the distribution of the natural logarithm of the
$D^0$ candidate's IP in Fig.~\ref{d0lnip_ss}.
Both Prompt and Dfb components are visible.  The IP for the Prompt signal would be zero without the effects of resolution. The Prompt shape is described by  a bifurcated double-Gaussian function. The distribution for Dfb is widely spread as the finite $b$ lifetime causes the $D^0$ meson not to point to the primary vertex;  we use a Monte-Carlo simulated shape.  The histogram in Fig.~\ref{d0lnip_ss} shows the results of a fit to the two components, letting the parameters of the Prompt shape float; this shape is used in systematic studies.

\section{\boldmath Evaluation of the $b\to D^0 X\mu^-\overline{\nu}$ yields}
\subsection{Using microbias data}
To select the decay chain $b\to D^0 X\mu^-\nu$,  $D^0\to K^-\pi^+$ and enrich our $b$ sample, we match $D^0$ candidates with tracks identified as muons, by ensuring that they penetrate the iron of the MUON system and have minimum ionization in the calorimeters \cite{LHCb-det}. Right-sign (RS) combinations have the sign of the charge of the muon being the same as the charge of the kaon in the $D^0$ decay. Wrong-sign (WS) combinations have the signs of the charges of the kaon and the muon being opposite; they are highly suppressed in semileptonic $b$ decay. WS events are useful to estimate certain backgrounds.

To find $b$ candidates we select $D^0$ candidates using the generic criteria specified above, and add a track that is identified as a muon, has $p_{\rm T}~>$ 500~MeV, and has $\chi^2_{\rm IP}>4$.
The $D^0$ and muon candidates are required to form a common
vertex with $\chi^2<5$, the $D^0\mu^-$ invariant mass must be between
3 and 5 GeV, and  the cosine of the angle of the $b$ pseudo-direction, formed from the $D^0$ and muon vector momentum sum with respect to the line between the $D^0\mu^-$ vertex and the primary vertex, must be $>$ 0.998. This angle cut is loose enough to have  about 97\% efficiency for  $b\to D^0 X\mu^- \overline{\nu}$  decays when taking into account the effect of the missing neutrino momentum.
We measure $\eta$ using the line defined by connecting the primary event vertex and the vertex formed by the $D^0$ and the $\mu^-$. Bins in $\eta$ are chosen to be larger than the  resolution  to obviate the need for any cross-feed corrections. Events are accepted in the interval $2<\eta<6$.

\begin{figure}[bt]
\centering
\includegraphics[width=3.in]{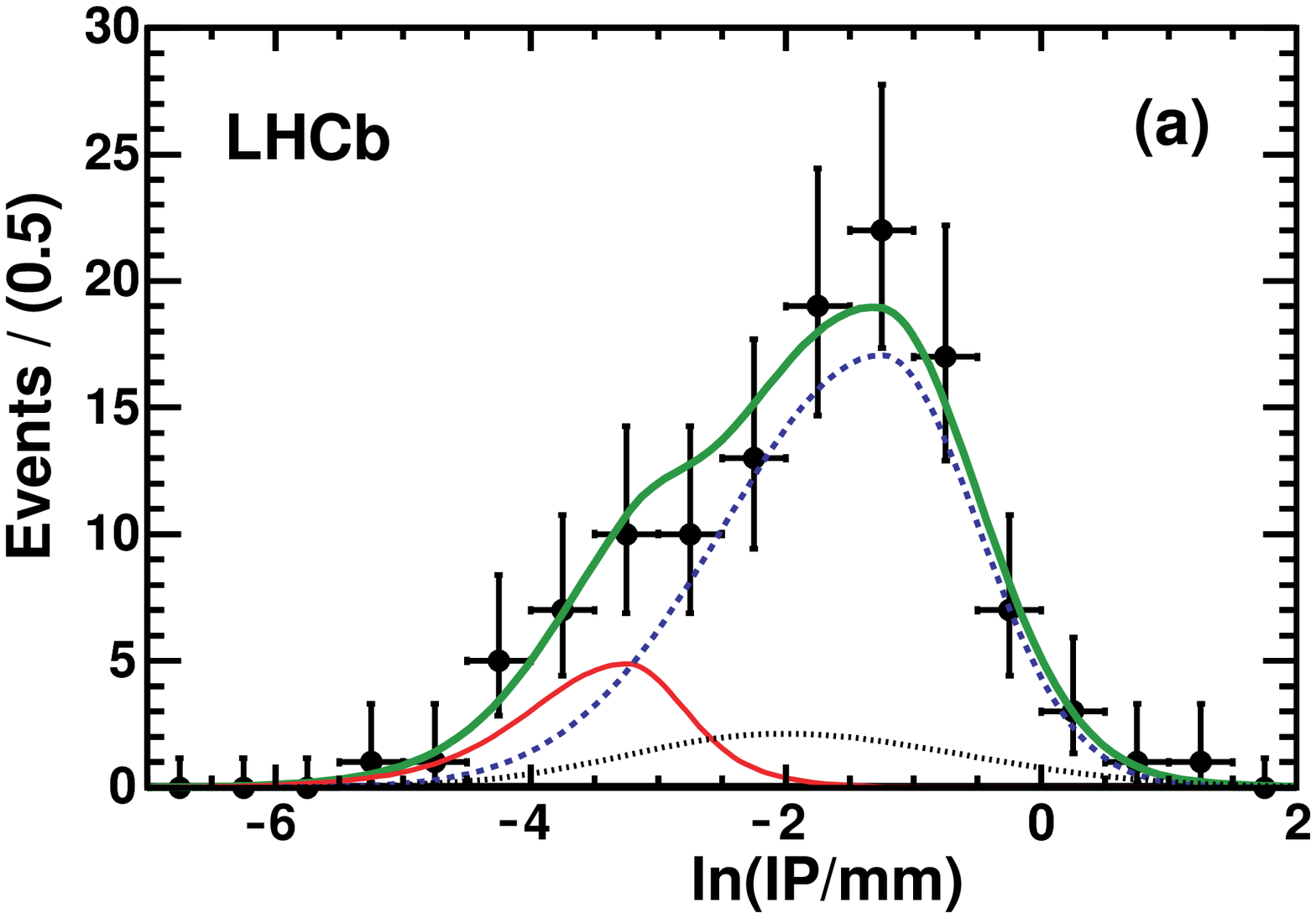}\hskip 0.02mm
\includegraphics[width=3.in]{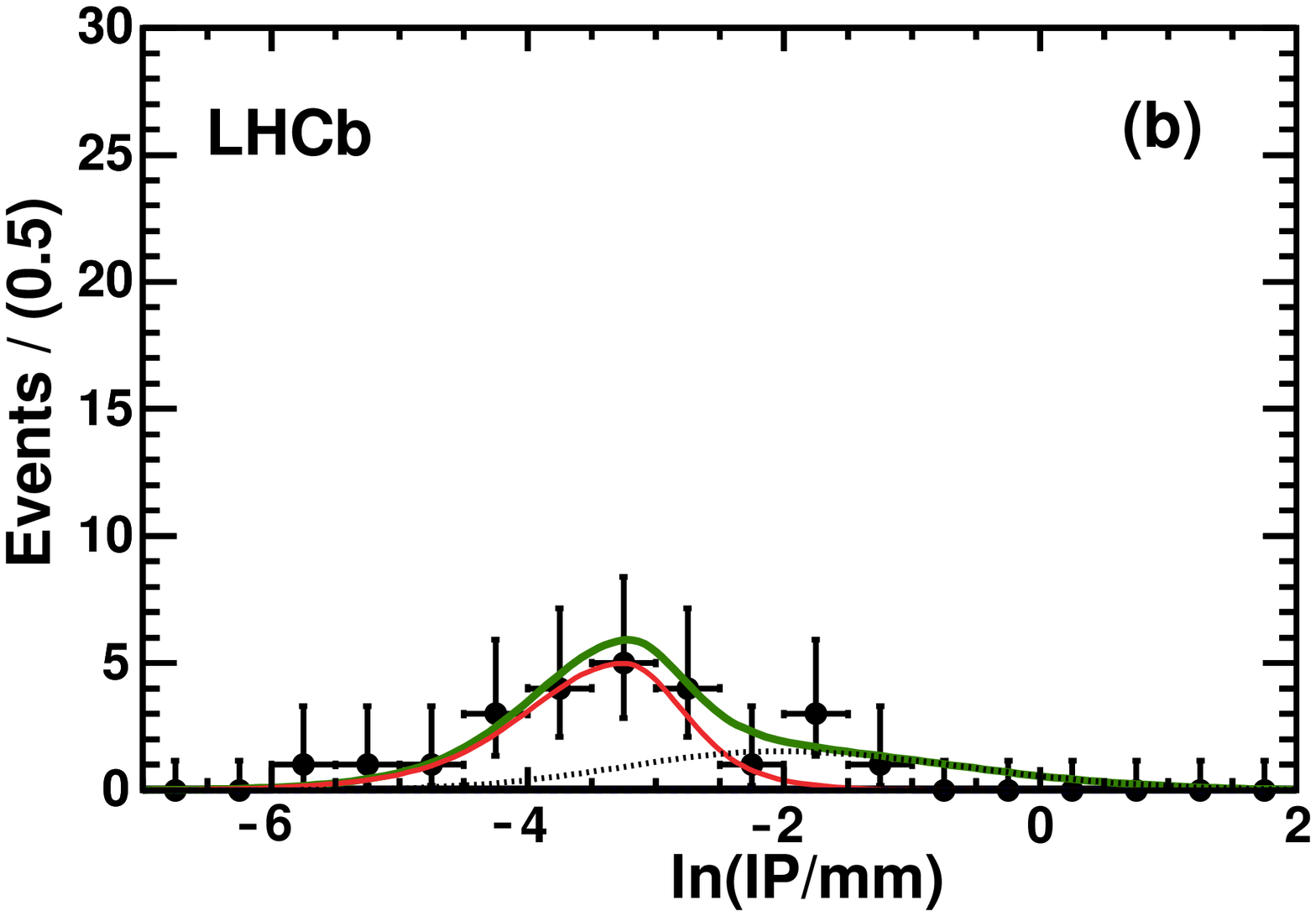}
\caption{\small Natural logarithm of the $D^0$ IP  in the 2.9 nb$^{-1}$ microbias sample for (a) right-sign and (b) wrong-sign $D^0$-muon candidate combinations. The dotted curves show the $D^0$ sideband backgrounds, the thin solid curves the Prompt yields, the
dashed curve  the Dfb signal, and the thick solid curves the totals.
 } \label{logip_bs}
\end{figure}

The IP distributions of both RS and WS candidates, requiring that the
$K^-\pi^+$ invariant mass is within 20 MeV of the $D^0$ mass, are shown in Fig.~\ref{logip_bs}. We perform an unbinned extended maximum likelihood fit to the two-dimensional distributions in $K^-\pi^+$ invariant mass over a region extending $\pm$100 MeV from the $D^0$ mass peak, and ln(IP/mm). This fitting procedure allows us directly to
determine the background shape from false combinations
under the $D^0$ signal mass peak. The parameters of the Prompt IP distribution are found by applying the
same criteria as for Fig.~\ref{logip_bs}, but with the additional track failing the muon identification criteria. The Monte Carlo simulated shape is used for the Dfb component.

The fit yields in the RS sample are 84.1$\pm$10.4 Dfb events, 16.3$\pm$5.4 Prompt events, and
14.0$\pm$1.9 background. In the WS the corresponding numbers are 0.0$\pm$1.1 Dfb events,
14.9$\pm$4.2 Prompt events, and
10.1$\pm$1.5 background. The Prompt yields are consistent between RS and WS as expected.

The contribution of tracks misidentified as muons (fakes) in both the RS and WS samples is evaluated by counting the number of tracks that satisfy all our criteria by forming a common vertex with a $D^0$ signal candidate, but do not satisfy our muon identification criteria. These tracks are categorized by their identity as electrons using ECAL, or pions, kaons or protons using the RICH.  These samples are then multiplied by the relevant fake rates that were estimated from simulation and checked with data.  The resulting ln(IP) distributions are examined, resulting in estimates of 2.2$\pm$0.4 RS Dfb fakes and 1.1$\pm$0.4 WS Dfb fakes.
The ${\cal {B}}(b\to D^0 X\tau^-\overline{\nu},~\tau^-\to\mu^-\nu\overline{\nu})$ of (0.36$\pm$0.11)\% is
 (5.3$\pm$1.6)\% of the semimuonic decay \cite{Lepers}. However, the relative efficiency to detect the resulting secondary muon is only 29\% leading to a 1.5\% subtraction. The lower efficiency is due to the lower secondary muon momentum from $\tau$ decay and the finite $\tau$ lifetime that causes some events to fail the vertex $\chi^2$ requirement. Other sources of backgrounds from $b$-hadron decays as evaluated by Monte Carlo simulation are small within our selection requirements, and predicted to be similar in size to the WS yields that are consistent with zero.

\subsection{Using muon triggered data}
\label{sec:HLT1selected}

\begin{figure}[bt]
\centering
\includegraphics[width=5.5in]{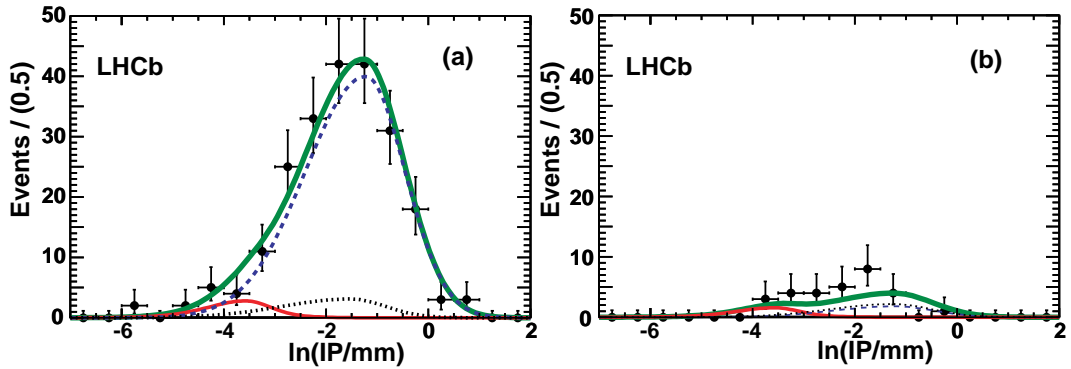}
\caption{\small
Natural logarithm of the $D^0$ IP  in the 12.2 nb$^{-1}$ triggered sample for (a) right-sign and (b) wrong-sign $D^0$-muon candidate combinations. The dotted curves show the $D^0$ sideband backgrounds, the thin solid curves the Prompt yields, the
dashed curve the Dfb signal, and the thick solid curves the totals.
 } \label{logip_11nb}
\end{figure}

The trigger imposes a cut of $p_{\rm T}>1.3$~GeV on muon candidates. The  IP distributions for both RS and WS
combinations are shown in Fig.~\ref{logip_11nb}.
We find a total of 195.4$\pm$14.9 RS Dfb, and 8.8$\pm$5.1 WS Dfb events. The Prompt contributions
are determined to be 9.3$\pm$4.8 RS with 5.3$\pm$3.0 WS.

In order to extract the $b$ cross-section from this data sample we have to make an additional correction for the overall  $\eta$-dependent trigger efficiency. The Monte Carlo simulated efficiency is checked using data by studying $J/\psi\to\mu^+\mu^-$ decays in
microbias events or those that triggered independently of the single muon trigger.
The data show a somewhat larger relative efficiency than the simulation, from 2\% at low $\eta$ rising to 11\% at
high $\eta$. We correct for this factor and use the 2\% error determined on the correction, to account for its uncertainty, that we add to the statistical error of this sample.

The IP distributions in each $\eta$ bin in both trigger samples are fit independently to the same functions as described above to extract the
$\eta$-dependent event yields.
The yields are listed in Table~\ref{tab:cross-section-HLT1}.
Muon fakes and the $\tau^-$ contribution are subtracted in the same manner as in the microbias sample.  In the triggered sample the
hadron-to-muon fake rates are smaller
as a result of the harder muon $p_{\rm T}$ cut imposed by the trigger of 1300~MeV rather than the 500 MeV used in analysing the microbias sample. The RS Dfb fakes total 1.0$\pm$0.2 and the WS Dfb fakes total 0.6$\pm$0.2 events. A uniform 1.5\% $\tau^-$ subtraction is done in each $\eta$ bin.

\begin{table}[hbt]
\begin{center}
\caption{\small RS background subtracted event yields from data, and extracted cross-sections, compared with predictions from  MCFM \cite{MCFM} and FONLL \cite{FONLL}. The systematic uncertainties in the normalisation of 17.3\% are not included. The uncertainties on the FONLL prediction are $^{+45}_ {-38}$\%, while those for MCFM are $^{+83}_ {-44}$\%.}\label{tab:cross-section-HLT1}
\begin{tabular}{c|cc|cccccc}\hline\hline
Bin& \multicolumn{2}{|c|}{Event yields}& \multicolumn{5}{c}{$\sigma(pp\to H_b X)$ ($\mu$b)}\\
 &Microbias& Trig.  & Microbias&Trig.& Average& {\footnotesize MCFM} & {\footnotesize FONLL}\\
 \hline
$2<\eta<3$& 16.7$\pm$4.5& 48.8$\pm$7.5& 27.2$\pm$7.3& 29.7$\pm$4.6 & 29.0$\pm$3.9 & 37.8  & 28.9 \\
$3<\eta<4$& 50.1$\pm$8.0& 111.4$\pm$11.0&28.8$\pm$4.6 &28.8$\pm$2.8& 28.8$\pm$2.4& 27.1 & 22.4\\
$4<\eta<5$&18.1$\pm$5.0& 30.2$\pm$6.0  &13.3$\pm$3.7 &11.7$\pm$2.3& 12.2$\pm$2.0 & 16.7 &13.1\\
$5<\eta<6$& ~4.7$\pm$2.8& ~5.2$\pm$2.2  &~6.5$\pm$3.6 &~4.8$\pm$2.5 &~5.3$\pm$2.0& ~7.4 &~5.9\\\hline
Sum&89.6$\pm$10.8 &195.6$\pm$14.9 &75.9$\pm$10.0 & 75.0$\pm$6.5& 75.3$\pm$5.4& 89.0 & 70.2\\
\hline\hline
\end{tabular}
\end{center}
\end{table}

\section{Luminosity determination and systematic uncertainties}
The luminosity was measured at specific periods during the data taking using
both Van der Meer scans and the `beam-profile' method \cite{lumin}.  Two Van der Meer scans were performed in a single fill. Analysis of these scans yielded consistent results
for the absolute luminosity scale with a precision of around 10\%, dominated by the
uncertainty in the knowledge of the beam currents.  In the second approach, six separate periods of stable running were chosen, and the beam-profiles measured using beam-gas and beam-beam interactions. Using these results, correcting for crossing angle effects, and knowing the beam currents, we determine the luminosity in each period following the analysis procedure described in Ref.~\cite{lumin}.  Consistent results were found for the absolute luminosity scale in each period, with a precision of 10\%, again dominated by the beam current uncertainty. These results are in good agreement with those of the Van der Meer analysis.

The knowledge of the absolute luminosity scale was used to calibrate the number of VELO tracks reconstructed using only the R sensors \cite{LHCb-det}, which are found to have a stable response throughout the data-taking period.
The integrated luminosities of the runs considered in this analysis were determined to be (2.85$\pm$0.29) and (12.2$\pm$1.2)~$\rm nb^{-1}$, respectively, for the microbias and triggered samples.

The product of detector acceptance, tracking efficiencies and our analysis cuts, as estimated by Monte Carlo simulation, is about 8\% for $b$ hadrons produced in the region $2<\eta<6$.
The systematic uncertainty on the tracking efficiency is evaluated by comparing the ratio of $D^0\to K^-\pi^+\pi^+\pi^-$ to $D^0\to K^-\pi^+$ events in data to the ratio in simulation. We find that
the ratio of data to Monte Carlo  efficiencies is 1.00$\pm$0.03 for tracks from $D^0$ decay, and use 3\% as the uncertainty per track.
For  the higher momentum muon track 4\% is used.
The total tracking uncertainty then being fully correlated is taken as 10\%, where this uncertainty is dominated by the
size of the data sample.
The kaon and pion RICH identification efficiencies are determined in each $\eta$ bin from a comparison of $D^0\to K^-\pi^+$ yields evaluated both with and without kaon identification.  An error of 1.5\% is set on the particle identification efficiencies that is mostly due to the kaon, as the pion selection criteria are much looser.

The efficiency of our muon selection criteria with respect to that obtained from the Monte Carlo simulation is evaluated as a function of momentum by detecting $J/\psi\to\mu^+\mu^-$ decays where one muon is  identified by passing our muon identification criteria while the opposite-sign track must  have been biased neither by the muon trigger, nor the muon identification criteria.
Using the momentum weighted averages we find (data/MC)  = $(96.9^{+2.4}_{-2.5})$\%. We correct for the difference and assign a 2.5\% error to our muon identification.

Since the $b\to D^0X\mu^-\overline{\nu}$ detection efficiency changes linearly with $p_{\rm T}$ by about a factor of four from zero to 12 GeV and then stays flat, the efficiency will not be estimated correctly if the Monte Carlo generator does not accurately simulate the $p_{\rm T}$ distribution. We investigate this possible efficiency change by examining the difference between the measured and simulated summed $p_{\rm T}$ distribution of the $D^0$ plus muon. They are consistent, and an uncertainty of 3\% is assigned as the systematic error from considerations of how large a difference the data allow.

Because the detection efficiency is different for $D^0$ mesons that result from $B^-\to D^{(*)0}\mu^-\overline{\nu}$
compared to those from other $b$ decays (such as $\overline{B}^0\to D^{*+}\mu^-\overline{\nu}$, $B\to D^{**}\mu^-\overline{\nu}$, $\overline{B}_s^0\to D_s^{**}\mu^-\overline{\nu}$, or similarly from $\Lambda_b$), we include an uncertainty due to the precision of our knowledge of the branching fractions \cite{PDG}. By varying these rates within their errors, we find an uncertainty of 4.4\%.  As discussed below, to translate our results on the yields into cross-section measurements we assume the fractions for fragmentation into the different $b$-hadron species as measured by LEP.  Varying these values within their errors gives a systematic uncertainty of $4.2\%$.

%Because the detection efficiency changes for $D^0$ mesons that result from $B^-\to D^{(*)0}\mu^-\overline{\nu}$ versus $B\to %D^{**}\mu^-\overline{\nu}$ decays, and $B_s^0$ produce $D^0$ mesons via $D_s^{**}$, or a related fragmentation process (similar for %$\Lambda_b$), we include an uncertainty due to possible changes in the mixture of these final states. By varying these rates within their errors, %we find an uncertainty of 4.4\%.

The efficiency of the various selection criteria with respect to simulation has been evaluated by changing the cuts. The resulting changes of the yield are small. The $D^0\mu^-$ vertex $\chi^2$ cut efficiency was cross-checked comparing data and Monte Carlo using  $\Xi^-\to\Lambda\pi^-$ decays. All of the uncertainties considered are listed in Table~\ref{tab:BBrsys}. The total systematic
uncertainty due to all sources added in quadrature is 17.3\%.

\begin{table}
\begin{center}
\caption{\small Systematic uncertainties.}\label{tab:BBrsys}
\begin{tabular}{lc|lc}\hline\hline
Source & Error (\%)&Source & Error (\%)\\\hline
Luminosity & 10.0&Prompt \& Dfb shapes & 1.4\\
Tracking efficiency   &  10.0  &    ${\cal{B}}\left(D^0\to K^-\pi^+\right)$ & 1.3 \\
${\cal{B}}(b\to D^0 X\mu^-\overline{\nu})$ & 5.1 &$D^0\mu^-$ vertex $\chi^2$ cut& 1.2\\
Assumed branching fractions & 4.4 &Kaon identification & 1.2\\
LEP fragmentation fractions & 4.2 &Muon fakes & 1.0\\
Generated $b$ $p_{\rm T}$ distribution &3.0&$D^0$ mass cut & 1.0\\
Muon identification & 2.5&$D^0$ vertex $\chi^2$ cut & 0.6\\
$\chi^2_{\rm IP}$  cut& 2.5&$D^0$ flight distance cut & 0.4\\
MC statistics & 1.5  &Pion identification & 0.3\\
\hline
\multicolumn{2}{c}{Total}&\multicolumn{2}{c}{17.3\%}\\
\hline \hline
\end{tabular}
\end{center}
\end{table}

\section{Cross-sections and comparison with theory}
\begin{figure}[bt]
\centering
\includegraphics[width=4.4in]{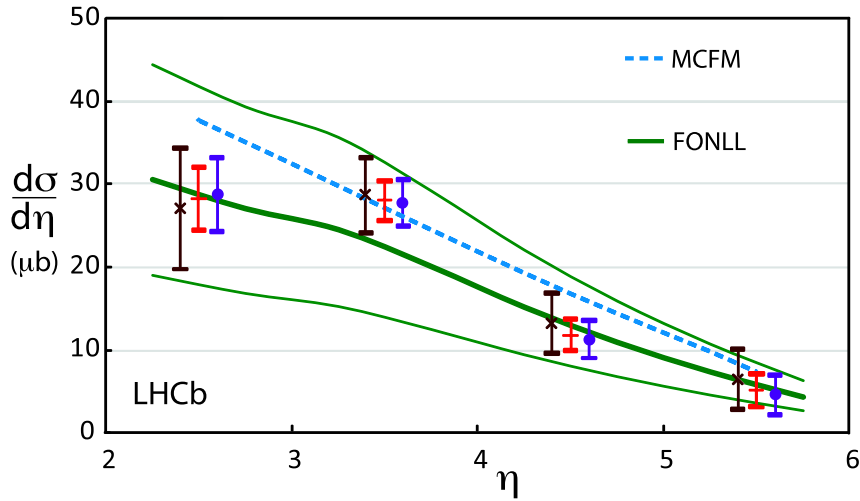}\vskip -0.02mm
\caption{\small $\sigma(pp\to H_b X)$  as a function of $\eta$ for the microbias ($\times$) and triggered ($\bullet$) samples, shown displaced from the bin center and the average (+). The data are shown as points with error bars, the MCFM prediction  as a dashed line, and the FONLL prediction as a thick solid line. The thin upper and lower lines indicate the theoretical uncertainties on the FONLL prediction. The systematic uncertainties in the data are not included.
 } \label{cross-section-all}
\end{figure}

The extracted cross-sections are listed in Table~\ref{tab:cross-section-HLT1}. The $\eta$-dependent cross-section is shown in Fig.~\ref{cross-section-all} for both data sets and the average.  The agreement between the two data sets is excellent.

We compare with two theories that predict $b$ production cross-sections as a function of $\eta$.  MCFM \cite{MCFM} predicts the cross-section for $b\overline{b}$ quark production in next to leading order (NLO) using the MSTW8NL parton distribution function (PDF).
The FONLL \cite{FONLL} prediction uses the CTEQ6.5 PDF, and improves the NLO result with the resummation of $p_{\rm T}$ logarithms up to next-to-leading order. It also includes the b-quark fragmentation into hadrons.
The measured yields are averaged over $b$-flavoured and $\overline{b}$-flavoured hadrons, $H_b$, in $\eta$ intervals:
\begin{equation}
\sigma(pp\to H_b X)=\frac{{\rm \#~of~detected~}D^0\mu^-~{\rm and}~\overline{D}^0\mu^+~
{\rm events}}{2{\cal {L}}\times{\rm efficiency}\times{\cal{B}}(b\to D^0 X\mu^-\overline{\nu}){\cal {B}}\left(D^0\to K^-\pi^+\right)}.
\end{equation}

Averaging the cross-sections from both samples, and summing over
 $\eta$, we measure
%\begin{equation}
%  \sigma_{H_b/2}(pp\to b\overline{b} X) = (74.9\pm 5.3\pm 12.9)~\mu{\rm b}
%\end{equation}

 \begin{equation}
  \sigma(pp\to H_b X) = (75.3\pm5.4\pm13.0)~\mu{\rm b}
\end{equation}
in the interval $2<\eta<6$. The first error is statistical, the second systematic. The LEP fragmentation fractions are used for our central values \cite{CDF-frac}.  Use of these fractions provides internal consistency to our results as ${\cal{B}}(b\to D^0 X \mu^-\overline{\nu})$ was also
measured at LEP.
The measured value changes if the $b$-hadron fractions differ. Fractions have also been measured at the Tevatron, albeit with large uncertainties \cite{CDF-frac}. The largest change with respect to LEP is that the $b$-baryon percentage rises from (9.1$\pm$1.5)\% to (21.4$\pm$6.8)\%. If  the Tevatron fractions are used,  our result changes to (89.6$\pm$6.4$\pm$15.5)~$\mu{\rm b}$.

\section{Conclusions}
The cross-section to produce $b$-flavoured hadrons is measured to be
%\begin{equation}
%  \sigma_{H_b/2}(pp\to b\overline{b} X) =(74.9\pm 5.3 \pm 12.3.0)~\mu{\rm b}
%\end{equation}
\begin{equation}
  \sigma(pp\to H_b X) =(75.3\pm5.4\pm13.0)~\mu{\rm b}
\end{equation}
in the pseudorapidity interval $2<\eta<6$
over the entire range of $p_{\rm T}$ assuming the LEP fractions for fragmentation into $b$-flavoured hadrons.
%Using the Tevatron fractions the result changes to $(89.6\pm 6.4 \pm 15.5)~\mu{\rm b}$.
For extrapolation to the full $\eta$ region,  theories predict factors of 3.73 (MCFM), and 3.61 (FONLL), while PYTHIA 6.4 gives 3.77. Using a factor of 3.77 for our extrapolation, we find a total $b\overline{b}$ cross-section of
 \begin{equation}
 \sigma(pp\to b\overline{b} X)=(284\pm20\pm49)~\mu{\rm b}
 \end{equation}
based on the LEP fragmentation results; using the Tevatron fragmentation fractions the result increases by 19\%. The quoted
systematic uncertainty does not include any contribution relating to the extrapolation over the $\eta$ range where LHCb has no sensitivity.

The production of $b$-flavoured hadrons has been measured in $p\overline{p}$ collisions in 1.8 and 1.96 TeV collisions at the Tevatron.
The earlier measurements at 1.8 TeV appeared to be higher than the NLO theoretical predictions \cite{Run1}. More recent measurements by the CDF collaboration at 1.96 TeV are consistent with the NLO theory \cite{Run2}. The history has been reviewed
by Mangano \cite{Mangano-hist}.
Here, with a large energy increase to 7 TeV, we find that the measured cross-section is consistent with theoretical predictions, both in normalization and $\eta$-dependent shape.

\section*{Acknowledgments}
We express our gratitude to our colleagues in the CERN accelerator departments for the excellent performance of the LHC.
We thank the technical and administrative staff at CERN and at the LHCb institutes, and acknowledge support from the National Agencies: CAPES, CNPq, FAPERJ and FINEP (Brazil); CERN; NSFC (China); CNRS/IN2P3 (France); BMBF, DFG, HGF and MPG (Germany); SFI (Ireland); INFN (Italy); FOM and NWO (Netherlands); SCSR (Poland); ANCS (Romania); MinES of Russia and Rosatom (Russia); MICINN, XUNGAL and GENCAT (Spain); SNSF and SER (Switzerland); NAS Ukraine (Ukraine); STFC (United Kingdom); NSF (USA). We also acknowledge the support received from the ERC under FP7 and the R\'egion Auvergne.

\end{document}